\documentclass[%
 reprint,
superscriptaddress,
 amsmath,amssymb,
 aps,
pra,
]{revtex4-2}

\usepackage{graphicx}% Include figure files
\usepackage{dcolumn}% Align table columns on decimal point
\usepackage{bm}% bold math
\usepackage[unicode]{hyperref}

\newcommand{\beq}{\begin{equation}}
\newcommand{\eeq}{\end{equation}}

\newcommand{\vect}[1]{\mathbf{#1}}

\DeclareMathOperator{\im}{Im}

\newcommand{\nn}{\nonumber}

\newcommand{\Eq}[1]{Eq.~\eqref{#1}}

\def\rr{\mathbf{r}}

\begin{document}

\preprint{AIP/123-QED}

\title{Three-Body Scattering Hypervolume \\of Particles with Unequal Masses}% Force line breaks with \\

\author{Zipeng Wang}
% \altaffiliation[Also at ]{Peking University}%Lines break automatically or can be forced with \\
\author{Shina Tan}%
 \email{shinatan@pku.edu.cn}
\affiliation{ 
International Center for Quantum Materials, Peking University, Beijing 100871, China%\\This line break forced with \textbackslash\textbackslash
}%

%\collaboration{}%\noaffiliation

\date{\today}% It is always \today, today,

\begin{abstract}

We analyze the collision of three particles with arbitrary
mass ratio at zero collision energy, assuming arbitrary short-range potentials,
and generalize the three-body scattering hypervolume $D$
first defined for identical bosons in 2008. We solve the three-body Schr\"{o}dinger equation asymptotically when the three particles are far apart or one pair and a third
particle are far apart, deriving two asymptotic expansions of the wave function,
and the parameter $D$ appears at the order $1/B^4$, where $B$ is the overall size of the
triangle formed by the particles. We then analyze the ground state energy
of three such particles with vanishing or negligible two-body scattering lengths in a large periodic volume of side length $L$, where
the three-body parameter contributes a term of the order $D/L^6$. From this result
we derive some properties of a two-component
Bose gas with negligible two-body scattering lengths: its energy density at zero temperature, the corresponding generalized Gross-Pitaevskii equation, the conditions for the stability of the two-component mixture against collapse or phase separation,
and the decay rates of particle densities due to three-body recombination.

%\begin{description}
%You may use the \texttt{description} environment to structure your %abstract;
%use the optional argument of the \verb+\item+ command to give the %category of each item. 
%\end{description}
\end{abstract}

\maketitle

%\tableofcontents

\section{\label{sec:level1}INTRODUCTION}
How do particles, composite or fundamental, such as atoms, molecules, ions, atomic nuclei,
neutrons, protons, electrons, mesons, etc, interact at small collision energies? It depends on their electric charges.
If at least one of two such particles is electrically neutral, usually the effective interaction between the two particles
is dominated by the $s$-wave scattering length $a$, for collision energies that are so small
that the de Broglie wave length of each particle in the center-of-mass frame is much longer than
the physical range of the interaction. If we need more precise knowledge of the effective pairwise interaction
at small collision energies, we need to also know other parameters such as the $s$-wave effective range $r_s$,
the $p$-wave scattering volume $a_p$, etc. All these parameters can be extracted from the wave functions
for the two-body collision at collision energies equal to or close to zero, outside of the physical range of 
interaction. The $s$-wave scattering length $a$, for example, can be extracted from
the wave function $\phi(\vect s)$ of the two particles colliding at zero incoming kinetic energy and zero orbital angular momentum \cite{pethick2008bose,pitaevskii2016bose}:
\beq\label{phi2}
\phi(\vect s)=1-\frac{a}{s},~~\text{if }s>r_e,
\eeq
where $\vect s$ is the spatial vector extending from one particle to the other, and
$r_e$ is the range of the microscopic interaction.
The scattering length $a$ is a key parameter in the quantum few-body and many-body physics for particles
with small collision energies.

If we want to gain more precise knowledge of the effective interaction strengths of low energy particles, we
need to also study the wave function for the collision of three particles at zero incoming kinetic energy. The three-body Schr\"{o}dinger equation is usually not analytically solvable, even outside of the range of the microscopic interactions. But, in a prior paper by one of the present authors, the three-body Schr\"{o}dinger equation was solved asymptotically for the collision of three identical bosons at zero incoming kinetic energy and zero orbital angular momentum, yielding well-controlled
expansions of the three-body wave function $\Phi^{(3)}$ when three or two pairwise distances are large \cite{tan2008three}. In such expansions,
a three-body parameter named three-body scattering hypervolume $D$ emerges \cite{tan2008three}. When all three pairwise distances
go to infinity simultaneously for a fixed ratio of pairwise distances, the three-body wave function
has the following expansion, which may be called ``111-expansion" (since each particle is alone when they are
all far apart from each other):
\beq
\Phi^{(3)}=1-\Xi-\frac{\sqrt{3}\,D}{8\pi^3B^4}+O(B^{-5}\ln B),
\eeq
where $B=\sqrt{(s_1^2+s_2^2+s_3^2)/2}$ is the hyperradius of the triangle formed by the particles,
\beq
\vect s_1\equiv\vect r_2-\vect r_3,~~\vect s_2\equiv\vect r_3-\vect r_1,~~\vect s_3\equiv\vect r_1-\vect r_2,
\eeq
$\vect r_i$ is the position vector of the $i$th particle, and
 $\Xi$ is a sum of a few terms due to a typically nonzero two-body scattering length.
 If $a=0$, $\Xi=0$. When two particles are kept at a fixed distance but the third particle is far away from the two,
there is another expansion which may be called the ``21-expansion" (since two particles are held at a fixed distance but the remaining particle is far apart):
\beq
\Phi^{(3)}=\sum_{q=0}^\infty S^{(-q)},
\eeq
where $S^{(-q)}$ scales as $R^{-q}$ for $q\le 3$,
and scales as $R^{-q}$ times some polynomial of $\ln R$ for $q\ge4$, and $R$ is the distance between the center of mass of the two particles and the remaining particle. The function $S^{(-q)}$ is expressed in terms of the ``two-body special functions" such as the $\phi(\vect s)$ in \Eq{phi2}, with coefficients that depend on $R$.
The three-body scattering hypervolume $D$ appears at the order $R^{-4}$.

The scattering hypervolume $D$ is the three-body analog of the two-body scattering length $a$.  
It is a fundamental parameter determining the effective strength of three-body interactions
at small collision energies, if the microscopic interactions vanish or become negligible beyond a certain range.
For given microscopic interactions, one can numerically solve the Schr\"{o}dinger equation for the zero-energy collision
of the three particles and match the solution to either the 111-expansion or the 21-expansion to determine $D$ numerically.
$D$ have been numerically computed in this way for identical bosons interacting with hard-sphere \cite{tan2008three}, Gaussian \cite{zhu2017three}, square-well \cite{mestrom2019scattering} and Lennard-Jones \cite{mestrom2020van} potentials.

If the collision of the three particles is purely elastic, $D$ is a real number.
But if the two-body forces are so strongly attractive that they support two-body bound states,
as is the case for most neutral atoms, then the three-body collisions are usually not purely elastic:
two such particles may fall into one of the bound states, releasing the binding energy in the form
of the center-of-mass kinetic energy of the two and the kinetic energy of the remaining free particle;
these inelastic processes are called ``three-body recombination" in cold atoms physics \cite{moerdijk1996decay,fedichev1996three,esry1996role,esry1999recombination,nielsen1999low,bedaque2000three,braaten2001three,HammerRevModPhys.85.197}.
When there is three-body recombination, $D$ becomes complex, and
the three-body recombination rate constant is proportional to the imaginary part of $D$ \cite{zhu2017three,braaten2006universality}.

The three-body scattering hypervolume \cite{tan2008three} determines the effective three-body coupling constant in the effective-field theoretical description of low energy particles \cite{tan2008three,braaten1999quantum,HammerRevModPhys.85.197}. It is also directly related to the three-body parameters in three-meson systems \cite{BeanePhysRevLett.100.082004,BeanePhysRevD.76.074507}.

The three-body scattering hypervolume provides a three-body effective interaction which, if repulsive, can stabilize dilute quantum droplets \cite{bulgac2002dilute,bedaque2003quantum,blakie2016properties}.

In Section~\ref{sec:asymp} of this paper, we generalize the 111-expansion and the 21-expansion to the collision of three particles with \emph{unequal} masses, $m_1$, $m_2$, $m_3$, for which we find that the expansions take much more complicated forms. We assume that two or three of the colliding particles are electrically neutral,
and our expansions are also applicable to the collision of two neutral particles and one charged particle,
such as two neutral mesons and one charged particle, as long as the neutral particles are not significantly electrically polarizable in the field of the charged particle.
Although in a prior paper \cite{tan2008three} the two expansions were carried out to the order $B^{-7}$ and $R^{-7}$ respectively, in this paper we will only expand $\Phi^{(3)}$ to the order $B^{-4}$ and $R^{-4}$ respectively, the order at which the three-body scattering hypervolume first appears.
If two of the particles have equal mass but are \emph{not} spin-polarized identical fermions,
and the third particle has a different mass, the expansions we find in this paper are also applicable.
Our work is motivated by many cold atoms experiments in which two or three atomic species having different
atomic masses are mixed together. But we believe our work will also be of fundamental importance
for other research areas such as nuclear physics.

In Section~\ref{sec:box} we compute the ground state energy of three particles with unequal masses in a large periodic cubic volume, assuming vanishing or negligible two-body scattering lengths, such that the energy is dominated by the three-body scattering hypervolume $D$.

In Section~\ref{sec:bose-bose} we consider a zero-temperature Bose-Bose mixture having negligible two-body scattering lengths
and derive its energy density in terms of the three-body scattering hypervolumes $D_{1}$,
$D_{112}$, $D_{122}$, and $D_{2}$, and write down
the corresponding generalized Gross-Pitaevskii equation.
Here $D_{1}$ is the scattering hypervolume of three particles of type 1,
and $D_{112}$ is the scattering hypervolume of two particles of type 1 and one particle of type 2, and so on.
We then give the criteria for stability of the mixture against collapse or phase separation.
Finally we study the decay rates of the particle densities in such a mixture
due to three-body recombination, in a shallow trap.

\section{ASYMPTOTICS OF THE THREE-BODY WAVE FUNCTION\label{sec:asymp}}
Consider three particles, labeled 1, 2, and 3, having masses $m_1$, $m_2$, and $m_3$, respectively.
Suppose that they have interactions that are invariant under translation, rotation and Galilean transformations, and suppose the interactions vanish beyond a certain range.
Particles 1 and 2 have scattering length $a_3$, particles 2 and 3 have scattering length $a_1$,
and particles 3 and 1 have scattering length $a_2$.
If the three particles collide with zero energy and zero orbital angular momentum,
the three-body wave function $\Phi^{(3)}$ satisfies the Schr\"{o}dinger equation:
\begin{align}
&\Big[-\frac{\hbar^2}{2m_1}\nabla_1^2-\frac{\hbar^2}{2m_2}\nabla_2^2-\frac{\hbar^2}{2m_3}\nabla_3^2+V_1(s_1)+V_2(s_2)\nn\\
&\quad+V_3(s_3)+V_{123}(s_1,s_2,s_3)\Big]\Phi^{(3)}(\vect r_1,\vect r_2,\vect r_3)=0,\label{3body_equ}
\end{align}
where $\vect r_i$ is the position vector of particle $i$, and
\beq
\vect s_i\equiv\vect r_j-\vect r_k.
\eeq
In the above equation and in the rest of the paper:
\begin{align}
   &\text{if }i=1,\text{ then }j=2, ~k=3;\notag\\
   &\text{if }i=2,\text{ then }j=3, ~k=1;\notag\\
   &\text{if }i=3,\text{ then }j=1, ~k=2.\notag
\end{align}
$V_i(s_i)$ is the interaction potential between particles $j$ and $k$,
and $V_{123}(s_1,s_2,s_3)$ is the three-body potential.
Note that $V_i$ and $V_{123}$ are \emph{not} zero-range pseudopotentials.
They are real potentials that extend to some nonzero pairwise distances.
But in this paper we assume that these potentials vanish beyond a certain range.
$\Phi^{(3)}$ is translationally invariant because of the zero total linear momentum:
\beq
\Phi^{(3)}(\vect r_1+\delta\vect r,\vect r_2+\delta\vect r,\vect r_3+\delta\vect r)=
\Phi^{(3)}(\vect r_1,\vect r_2,\vect r_3)
\eeq
for all $\delta\vect r$. 
$\Phi^{(3)}$ is also rotationally invariant because of the zero orbital angular momentum.
As a result, $\Phi^{(3)}$ depends only on the pairwise distances $s_1$, $s_2$, and $s_3$.
We choose the amplitude of $\Phi^{(3)}$ such that it approaches $1$ when
the three pairwise distances all go to infinity.

\subsection{Jacobi coordinates, hyperradius, and hyperangles}
For later use, we define $\vect{R}_i$ as the vector extending from the center of mass of particles $j$ and $k$
to particle $i$:
\begin{equation}
\vect{R}_i\equiv\vect{r}_i-\frac{m_{j}\vect{r}_{j}+m_{k}\vect{r}_{k}}{m_{j}+m_{k}}.
\end{equation}
($\vect s_i$,$\vect R_i$) are called Jacobi coordinates \cite{braaten2006universality,nielsen2001three}.
We define the \textit{hyperradius} $B$ as
\begin{equation}
B\equiv\sqrt{\frac{3}{2}\frac{m_1m_2 s_3^2+m_2m_3s_1^2+m_3m_1s_2^2}{m_1m_2+m_2m_3+m_3m_1}}.
\end{equation}
If the particles have equal mass, the above definition of $B$ reduces to the one
in Ref.~\cite{tan2008three}.
Let $\mu_i$ be the reduced mass of particles $j$ and $k$, and $\nu_i$ be the reduced mass
of the particle $i$ and the pair $jk$:
\begin{equation}
\frac{1}{\mu_i}\equiv\frac{1}{m_{j}}+\frac{1}{m_{k}},\quad\frac{1}{\nu_i}\equiv\frac{1}{m_i}+\frac{1}{m_{j}+m_{k}}.
\end{equation}
They satisfy
\begin{equation}
    \mu_i \nu_i =\frac{m_1 m_2 m_3}{m_1+m_2+m_3}.
\end{equation}
Define
\begin{equation}
   \epsilon_i\equiv\sqrt{\frac{\mu_i}{\nu_i}},
\end{equation}
\begin{equation}
\lambda\equiv\frac{3}{2}\frac{m_1+m_2+m_3}{m_1 m_2+m_2 m_3+m_3 m_1}.
\end{equation}
One can show that
\begin{equation}
B^2=\lambda \left ( \nu_i R_i^2+\mu_i s_i^2\right)
\end{equation}
for any $i\in\{1,2,3\}$.
We also define three \textit{hyperangles}:
\begin{equation}
\theta_i\equiv\arctan\frac{R_i}{\epsilon_i s_i}.
\end{equation}
%The three $\theta_i$ are not independent, there's an identity:
%\begin{equation}
%\sum_{i}(m_{i+1}+m_{i+2})\cos2\theta_i=0.
%\end{equation}
$s_i, R_i$ and $B$ satisfy
\begin{equation}
s_i=\frac{1}{\sqrt{\lambda \mu_i}} B \cos\theta_i,\quad R_i=\frac{1}{\sqrt{\lambda \nu_i}} B \sin \theta_i.
\end{equation}

%In this paper, we will often use the subscript 1 to represent the quantity related to particle 2 and 3, and the subscript 2 is related to particle 3 and 1, and 3 is related to particle 1 and 3.In the following sections, $a_1=a_{23}$, $a_2=a_{31}$, $a_3=a_{12}$, where $a_{ij}$ is the $s$-wave scattering length between particle $i$ and $j$. It's similar for the $p$-wave scattering volume $a_{i,p}$, the effective range $r_{i,(l)}$, and the reduced mass $\mu_1=\mu_{23}$, $\mu_2=\mu_{31}$, $\mu_3=\mu_{12}$. Sometimes we also take $a_i=a_{i+3}$, $m_{i}=m_{i+3}$ in summation expression.

\subsection{\label{sec:level2}Two-body special functions}
For $i=1$, 2, or 3, we define the
two-body special functions $\phi_{i,\hat{\vect n}}^{(l)}(\vect s)$, $f_{i,\hat{\vect n}}^{(l)}(\vect s)$, $g_{i,\hat{\vect n}}^{(l)}(\vect s)$, \dots for the collision
of particles $j$ and $k$ with orbital angular momentum quantum number $l$ and zero magnetic quantum number along the direction specified by the unit vector $\hat{\vect n}$ \cite{tan2008three}:
\begin{equation}
    \widetilde{H}_i\phi_{i,\hat{\vect n}}^{(l)}=0,\quad \widetilde{H}_i f_{i,\hat{\vect n}}^{(l)}=\phi_{i,\hat{\vect n}}^{(l)},\quad \widetilde{H}_i g_{i,\hat{\vect n}}^{(l)}=f_{i,\hat{\vect n}}^{(l)},\dots
\end{equation}
where $\hbar^2\widetilde{H}_i/2\mu_i$ is the two-body Hamiltonian for the collision of particles $j$ and $k$
in the center-of-mass frame, and
\begin{equation}
    \widetilde{H}_i \equiv -\nabla_{\vect s}^2+ \frac{2\mu_i}{\hbar^2} V_i(s).
\end{equation}
Unlike the case of identical bosons \cite{tan2008three}, here $l$ may be odd.

To complete the definition of $\phi_{i,\hat{\vect n}}^{(l)}$, we need to specify its overall amplitude.
Since the potential $V_i(s)$ vanishes beyond a finite range $r_e$, $\phi_{i,\hat{\vect n}}^{(l)}$ takes a simple form at $s>r_e$:
\beq\label{two-body-phi}
\phi^{(l)}_{i,\hat{\mathbf{n}}}(\mathbf{s})=\left[ \frac{s^l}{(2l+1)!!}-\frac{(2l-1)!!a_{i,l}}{s^{l+1}}\right]P_l(\hat{\mathbf{n}}\cdot\hat{\mathbf{s}}),
\eeq
where we have fixed the overall amplitude of $\phi_{i,\hat{\vect n}}^{(l)}$ by specifying the coefficient of the term $\propto s^l$.
Here $P_l$ is the Legendre polynomial. The parameter $a_{i,l}$ is determined by solving the two-body Schr\"{o}dinger equation
at zero collision energy, $\widetilde{H}_i\phi_{i,\hat{\vect n}}^{(l)}=0$, using the two-body potential $V_i(s)$ at $s<r_e$,
and matching the solution with \Eq{two-body-phi} at $s>r_e$.

The solution to the equation $\widetilde{H}_i f_{i,\hat{\vect n}}^{(l)}=\phi_{i,\hat{\vect n}}^{(l)}$ is not unique, because if $f_{i,\hat{\vect n}}^{(l)}(s)$ satisfies this equation, then $f_{i,\hat{\vect n}}^{(l)}(s)+\text{(arbitrary coefficient)}\times\phi_{i,\hat{\vect n}}^{(l)}(s)$ also satisfies this equation.
To complete the definition of $f_{i,\hat{\vect n}}^{(l)}(s)$, we specify that in the expansion of $f_{i,\hat{\vect n}}^{(l)}(s)$ at $s>r_e$ we do \emph{not} have
the term $\propto s^{-l-1}$ (if such a term exists, we can add a suitable coefficient times $\phi_{i,\hat{\vect n}}^{(l)}(s)$ to $f_{i,\hat{\vect n}}^{(l)}(s)$ to cancel this term). Then at $s>r_e$ we have the following analytical formula for $f_{i,\hat{\vect n}}^{(l)}(s)$:
\beq\label{two-body-f}
\begin{split}
&   f^{(l)}_{i,\hat{\mathbf{n}}}(\vect{s})=\bigg[-\frac{s^{l+2}}{2(2l+3)!!}-\frac{a_{i,l}r_{i,l}s^l}{2(2l+1)!!}\\
&\mspace{11mu}\quad\quad\quad\quad-\frac{(2l-3)!!}{2}a_{i,l}s^{1-l}\bigg]P_l\left(\hat{ \mathbf{n}}\cdot\hat{\mathbf{s}} \right).
\end{split}
\eeq
This completes the definition of $f_{i,\hat{\vect n}}^{(l)}(s)$.
We can similarly fix the definitions of $g_{i,\hat{\vect n}}^{(l)}(s)$ etc. For brevity we do not show the formula for $g_{i,\hat{\vect n}}^{(l)}(\vect s)$
at $s>r_e$ as it is not explicitly used in this paper.

%\begin{equation}
%\begin{split}
%&\phi_i(\vect{s})=1-\frac{a_i}{s},\\
%&\phi^{(p)}_{i,\hat{\mathbf{n}}}(\mathbf{s})=\left( \frac{s}{3}-\frac{a_{i,p}}{s^2}\right)P_1(\hat{\mathbf{n}}\cdot\hat{\mathbf{s}}), \\
%&\phi^{(d)}_{i,\hat{\mathbf{n}}}(\mathbf{s})=\left( \frac{s^2}{15}-\frac{3a_{i,d}}{s^3}\right)P_2(\hat{\mathbf{n}}\cdot\hat{\mathbf{s}}),\\
%&\phi^{(f)}_{i,\hat{\mathbf{n}}}(\mathbf{s})=\left( \frac{s^3}{105}-\frac{15a_{i,f}}{s^4}\right)P_3(\hat{\mathbf{n}}\cdot\hat{\mathbf{s}}),\\
%&...
%\end{split}
%\end{equation}

Given the two-body special functions $\phi_{i,\hat{\vect n}}^{(l)}$, $f_{i,\hat{\vect n}}^{(l)}$, $g_{i,\hat{\vect n}}^{(l)}$, \dots, we can express the wave function
for the collision of particles $j$ and $k$ at any small nonzero energy $E=\hbar^2 k^2/2\mu_i$ as an infinite series in $k^2$:
\begin{equation}\label{two-body-nonzero-E:expand}
\phi_{i,\hat{\vect n}}^{(l,k)}(\vect s)=\phi_{i,\hat{\vect n}} ^{(l)}(\vect s)+k^2 f_{i,\hat{\vect n}}^{(l)}(\vect s)+ k^4 g_{i,\hat{\vect n}}^{(l)}(\vect s)+\cdots.
\end{equation}
It is easy to verify that this series satisfies the Schr\"{o}dinger equation at nonzero energy $E$, namely
\beq\label{two-body-nonzero-E:eq}
\widetilde{H}_i\phi_{i,\hat{\vect n}}^{(l,k)}(\vect s)=k^2\phi_{i,\hat{\vect n}}^{(l,k)}(\vect s).
\eeq
On the other hand, if $s>r_e$ then $V_i(s)=0$ and \Eq{two-body-nonzero-E:eq} can be solved analytically to yield
\beq\label{two-body-nonzero-E:s>re}
\phi_{i,\hat{\vect n}}^{(l,k)}(\vect s)=\alpha_{i,l}(k)\big[j_l(ks)\cot\delta_{i,l}-n_l(ks)\big]P_l(\hat{\vect n}\cdot\hat{\vect s}),
\eeq
where $j_l$ and $n_l$ are spherical Bessel functions of the first kind and the second kind, respectively,
$\delta_{i,l}$ is the scattering phase shift, and the overall coefficient $\alpha_{i,l}(k)$ is to be determined.
Comparing \Eq{two-body-nonzero-E:s>re} with \Eq{two-body-nonzero-E:expand}, and using the definitions of $\phi_{i,\hat{\vect n}}^{(l)}$ and $f_{i,\hat{\vect n}}^{(l)}$ etc, we find that
$\alpha_{i,l}(k)=-k^{l+1}a_{i,l}$ and
\begin{equation}\label{effective-range-expansion}
k^{2l+1}\cot \delta_{i,l}(k)=-\frac{1}{a_{i,l}}+\frac{1}{2}r_{i,l} k^2+O(k^4).
\end{equation}
Equation~\eqref{effective-range-expansion} is in fact the well-known effective range expansion \cite{hammer2010causality}.
We now see that $a_{i,l}$ which first appears in \Eq{two-body-phi} is the two-body $l$-wave scattering length
(or volume or hypervolume) of particles $j$ and $k$, and 
$r_{i,l}$ which first appears in \Eq{two-body-f} is the two-body $l$-wave effective range.

For $l=0$, we write the functions $\phi_{i,\hat{\vect n}}^{(0)}(\vect s)$, $f_{i,\hat{\vect n}}^{(0)}(\vect s)$, and $g_{i,\hat{\vect n}}^{(0)}(\vect s)$ simply as $\phi_i(\vect{s})$, $f_i(\vect{s})$, and $g_i(\vect{s})$. 
We use symbols $s$, $p$, $d$, $f$,\dots to represent $l=0,1,2,3,\dots$.
For later convenience we simply write the $s$-wave scattering length $a_{i,s}$ as $a_i$.
%The functions $f_i^{(l)}$ and $g_i^{(l)}$ etc 
%can also be determined at $s>r_e$:
%\begin{equation}
%f_i(\mathbf{s})= -\frac{s^2}{6}+\frac{a_is}{2}-\frac{a_i r_{i,s}}{2},
%&f^{(p)}_{\hat{\mathbf{n}}}(\mathbf{r})=\left( -\frac{1}{30}r^3-\frac{1}{6}a_p r_p r-\frac{1}{2}a_p\right) P_1\left(\hat{ \mathbf{n}}\cdot\hat{\mathbf{r}} \right) ,\\
%&f^{(d)}_{\hat{\mathbf{n}}}(\rr)=\left(-\frac{1}{210}r^4-\frac{1}{30}a_d r_d r^2-\frac{a_d}{2r}\right)P_2\left(\hat{ \mathbf{n}}\cdot\hat{\mathbf{r}} \right),\\
%\end{equation}

The two-body special functions will appear in the expansion of the three-body wave function $\Phi^{(3)}$
when two particles are held at a fixed distance and the third particle is far away from the two.

\subsection{Asymptotics of $\Phi^{(3)}$ at large distances}
When particle $i$ is far away from particles $j$ and $k$, but particles $j$ and $k$ are held at a fixed
distance $s_i$, the pairwise interaction potentials $V_{j}(s_j)$ and $V_{k}(s_k)$ and the three-body potential $V_{123}(s_1,s_2,s_3)$ vanish, and \Eq{3body_equ} is simplified as
\beq
\Big[-\nabla_{\vect s_i}^2+\frac{2\mu_i}{\hbar^2}V_i(s_i)-\frac{\mu_i}{\nu_i}\nabla_{\vect R_i}^2\Big]\Phi^{(3)}=0.\label{3body SE}
\eeq
The following partial-wave expansion is the formal solution to the above equation:
\begin{align}
\Phi^{(3)}=\sum_{l=0}^\infty\Big[&A^{(l)}_i(R_i)\phi^{(l)}_{i,\hat{\vect R}_i}(\vect s_i)
+B^{(l)}_i(R_i)f^{(l)}_{i,\hat{\vect R}_i}(\vect s_i)+\nn\\
&+C^{(l)}_i(R_i)g^{(l)}_{i,\hat{\vect R}_i}(\vect s_i)+\cdots
\Big],
\end{align}
where the function $A^{(l)}_i(R_i)$ has a well-controlled expansion at large $R_i$, and $B^{(l)}_i(R_i)$, $C^{(l)}_i(R_i)$, \dots\, satisfy
\begin{subequations}
\begin{align}
B^{(l)}_i(R_i)&=\frac{\mu_i}{\nu_i}\Big[\frac{1}{R_i^2}\frac{d}{dR_i}R_i^2\frac{d}{dR_i}-\frac{l(l+1)}{R_i^2}
\Big]A^{(l)}_{i}(R_i),\\
C^{(l)}_i(R_i)&=\frac{\mu_i}{\nu_i}\Big[\frac{1}{R_i^2}\frac{d}{dR_i}R_i^2\frac{d}{dR_i}-\frac{l(l+1)}{R_i^2}
\Big]B^{(l)}_{i}(R_i),
\end{align}
and so on.
\end{subequations}
We may also group the terms according to the powers of $1/R_i$:
\begin{equation}
    \Phi^{(3)}(\vect{r}_1,\vect{r}_2,\vect{r}_3)=\sum_{q=0}^{\infty}S_i^{(-q)}(\vect{R}_i,\vect{s}_i),\label{21-form}
\end{equation}
where $S_i^{(-q)}$ scales as $R_i^{-q}$
times some polynomial of $\ln R_i$ (such a logarithmic dependence on $R_i$ could be absent for a small $q$), and satisfies
\begin{equation}
\begin{split}
    &\widetilde{H}_i S_i^{(0)}=0,\\
    &\widetilde{H}_i S_i^{(-1)}=0,\\
    &\widetilde{H}_i S_i^{(-q)}=\frac{\mu_i}{\nu_i}\nabla_{\vect{R}_i}^2 S_i ^{(-q+2)}\quad (q\geq 2). 
\end{split}
\end{equation}
Equation~\eqref{21-form} is the 21-expansion.

When the three particles are all far apart from each other, such that $s_1$, $s_2$, $s_3$ go to
infinity simultaneously for any fixed ratio $s_1:s_2:s_3$, we may expand $\Phi^{(3)}$ in powers of $1/B$:
\begin{equation}
    \Phi^{(3)}(\vect{r}_1,\vect{r}_2,\vect{r}_3)=\sum_{p=0}^{\infty} T^{(-p)}(\rr_1,\rr_2,\rr_3),\label{111-form}
\end{equation}
where $T^{(-p)}$ scales as $B^{-p}$ times some polynomial of $\ln B$ (such a logarithmic dependence on $\ln B$ could be absent for a small $p$), and satisfies the free Schr\"{o}dinger equation:
\beq\label{free Schrodinger}
\Big(-\frac{\hbar^2}{2m_1}\nabla_1^2-\frac{\hbar^2}{2m_2}\nabla_2^2-\frac{\hbar^2}{2m_3}\nabla_3^2\Big)
T^{(-p)}=0
\eeq
if the pairwise distances $s_i$ are all nonzero.
Equation~\eqref{111-form} is the 111-expansion.

We start from the leading-order term in the 111-expansion:
\beq
T^{(0)}=1,
\eeq
and first derive $S_i^{(0)}$, and then derive $T^{(-1)}$, and then derive $S_i^{(-1)}$, and then derive $T^{(-2)}$,
and so on, all the way until $S_i^{(-4)}$.
At every step, we require the 111-expansion and the 21-expansion to be consistent in the region
$r_e\ll s_i\ll R_i$. See the Appendix %\ref{app-1}
for details. Our resultant
111-expansion is
\begin{widetext}
\begin{equation}
\begin{split}
\Phi^{(3)}=&1-\frac{\sqrt{3}D}{8\pi^3 B^4}+\sum_{i=1}^{3}\bigg\{-\frac{a_i}{s_i}+\frac{2b_i\theta_i}{\pi R_i s_i}-\frac{2\lambda}{\pi} \frac{\nu_i w_i a_i}{B^2 s_i} +\frac{m_s}{B^4}
J_i^{(s)}\left[ \Big(\ln \frac{B}{\sqrt{\lambda \nu_i}|a_i|}\Big)+\gamma-1-\theta_i \cot( 2\theta_i)\right]\\
&+\frac{d_p}{B^4}J_i^{(p)}\frac{\sin (4\theta_i)-4\theta_i}{\sin^2 (2\theta_i)} (\hat{\vect{R}}_i\cdot\hat{\vect{s}}_i)\bigg\} +O(B^{-5}\ln^n B),
\end{split}\label{1-1-1 expansion}
\end{equation}
\end{widetext}
where $\gamma=0.57721566...$ is Euler's constant, $n$ is a nonnegative integer (we conjecture here $n=1$, just like the case of identical bosons \cite{tan2008three}), and
\begin{subequations}
\begin{equation}
    b_i=a_i (a_{j}+a_{k}),
\end{equation}
\begin{equation}
    w_i=-\epsilon_i b_i+\frac{\beta_{k}b_{j}}{\eta_{ik}}+\frac{\beta_{j}b_{k}}{\eta_{ij}},
\end{equation}
\begin{equation}
    \eta_{\mu\nu}=\frac{m_\mu}{m_\mu+m_\nu},~\text{for }\mu,\nu\in\{1,2,3\},
\end{equation}
\begin{equation}
    \beta_i=\arctan\sqrt{\frac{m_{j}m_{k}}{m_i(m_1+m_2+m_3)}},
\end{equation}
\begin{equation}
    m_s=\frac{18(m_1m_2m_3)^{3/2}\sqrt{m_1+m_2+m_3}}{\pi^2 (m_1m_2+m_2m_3+m_3m_1)^2},
\end{equation}
\begin{equation}
    d_p=\frac{27m_1m_2m_3(m_1+m_2+m_3)}{2 \pi (m_1m_2+m_2m_3+m_3m_1)^2},
\end{equation}
\begin{equation}
    J_i^{(s)}=a_i\left(\frac{w_j a_j}{\mu_j}+\frac{w_k a_k}{\mu_k}\right),
\end{equation}
\begin{equation}
    J_i^{(p)}=a_{i,p}\frac{m_j a_j-m_k a_k}{m_j+m_k}.
\end{equation}
\end{subequations}
$D$ is the \textit{three-body scattering hypervolume}, and its dimension is $[\mathrm{length}]^4$. It is the generalization of the scattering hypervolume for identical bosons defined in Ref.~\cite{tan2008three}.
The value and sign of the scattering hypervolume in each three-particle system depend on the details of two-body and three-body potentials, as well as the masses of the three particles.
For very weakly repulsive potentials $D$ is small and positive. For very weakly attractive potentials $D$ is small and negative. As one increases the strength
of attractive interactions such that they nearly support a three-body bound state, $D$ becomes large and negative. At the critical attraction strength at which the three-body $s$-wave bound state energy is zero, $D$ is divergent. As one slightly increases the strength of attraction further, then $D$ becomes large and positive. Further increasing the strength of attraction, one can make $D$ smaller. $D$ will pass zero and turn negative as one further increases the strength of attraction.
If the two-body potentials are sufficiently strongly attractive
such that there are two-body bound states, $D$ will in general acquire some imaginary part
which determines the three-body recombination rate constant for dilute gases consisting of
the relevant particles (the rate constant is proportional to $\im D$) \cite{zhu2017three}.
If there are multiple two-body bound states then $\im D$ contains the contributions from
the three-body recombination processes to all these two-body bound states.

%\begin{equation}
%D=\tilde{D}\frac{9(m_1m_2m_3)^{3/2}\sqrt{m_1+m_2+m_3}}{\sqrt{3}(m_1m_2+m_2m_3+m_3m_1)^2}.
%\end{equation}

Our resultant 21-expansion is
\begin{widetext}
\begin{equation}
\begin{split}
\Phi^{(3)}=&\bigg[ 1-\frac{a_{j}+a_{k}}{R_i}+\frac{2w_i}{\pi R_i^2}-\frac{2\mu_i}{\pi R_i^3}\left( \frac{w_{j}a_{j}}{\mu_{j}}+\frac{w_{k}a_{k}}{\mu_{k}}\right)+\frac{W_i}{R_i^4}+\frac{8\epsilon_i \mu_i J^{(s)}}{\pi^2 R_i^4}\ln \frac{R_i}{|a_i|}\bigg] \phi_i(\mathbf{s}_i)\\
&+\left(\frac{3}{R_i^2}\frac{m_{j}a_{j}-m_{k}a_{k}}{m_{j}+m_{k}}+\frac{3c_i}{R_i^3}+\frac{\#_{ip}}{R_i^4}\right) \phi_{i,\hat{\mathbf{R}}_i}^{(p)}(\mathbf{s}_i)+\left[ -\frac{15}{R_i^3}\frac{m_{j}^2a_{j}+m_{k}^2a_{k}}{(m_{j}+m_{k})^2}+\frac{\#_{id}}{ R_i^4}\right] \phi_{i,\hat{\mathbf{R}}_i}^{(d)}(\mathbf{s}_i)\\
&+\frac{105}{R_i^4}\frac{ m_{j}^3a_{j}-m_{k}^3a_{k}}{(m_{j}+m_{k})^3} \phi_{i,\hat{\mathbf{R}}_i}^{(f)}(\mathbf{s}_i)+\frac{4\epsilon_i^2 w_i}{\pi R_i^4}f_i(\mathbf{s}_i) +O(R_i^{-5}\ln^n R_i),
\end{split}\label{2-1 expansion}
\end{equation}
\end{widetext}
where $\phi_{i,\hat{\vect{R}}_i}^{(l)}$ and $f_i$ are the two-body special functions for particles $j$ and $k$ as defined in Sec. \ref{sec:level2}, and
\begin{subequations}
\begin{equation}
    J^{(s)}=J_1^{(s)}+J_2^{(s)}+J_3^{(s)},
\end{equation}
\begin{equation}
\begin{split}
c_i=-&\frac{2b_{j}}{\pi\eta_{ik}^2}[ \epsilon_j\eta_{jk}+\left( 2\eta_{jk}\eta_{ik}-1\right) \beta_{k}] \\
+&\frac{2b_{k}}{\pi\eta_{ij}^2}[ \epsilon_k \eta_{kj}+\left( 2\eta_{kj}\eta_{ij}-1\right) \beta_{j}] ,
\end{split}
\end{equation}

\begin{equation}
\#_{ip}=\frac{6\mu_i^2}{\pi}\left( \frac{w_{j}a_{j}}{\mu_{j} m_{k}}-\frac{w_{k}a_{k}}{\mu_{k}m_{j}}\right),
\end{equation}

\begin{equation}
\begin{split}
\#_{id}&=\frac{10}{\pi}\left\lbrace  3b_{j}\left[ \frac{\epsilon_{j}\mu_i}{\eta_{ik}^2 \mu_{j}}(2\eta_{jk}\eta_{ik}-1)\right.\right.\\
&\left.\left.+\frac{1}{\eta_{ik}^3}(1-2\eta_{jk}\eta_{ik}+2\eta_{jk}^2\eta_{ik}^2)\beta_{k}-\epsilon_i^2 \frac{2\beta_{k}}{3\eta_{ik}}\right] \right. \\
&\left.+3b_{k}\left[ \frac{\epsilon_{k}\mu_i}{\eta_{ij}^2 \mu_{k}}(2\eta_{kj}\eta_{ij}-1)\right.\right.\\
&\left.\left.+\frac{1}{\eta_{ij}^3}(1-2\eta_{kj}\eta_{ij}+2\eta_{kj}^2\eta_{ij}^2)\beta_{j}-\epsilon_i^2 \frac{2\beta_{j}}{3\eta_{ij}}\right] \right\rbrace ,
\end{split}
\end{equation}

\begin{equation}
\begin{split}
     &W_i=\\
     &-\frac{(m_1m_2+m_2m_3+m_3m_1)^2D}{6\sqrt{3}\pi^3 m_i^2(m_{j}+m_{k})^2}+\frac{2\epsilon_i^2 w_i a_i r_{i,s}}{\pi}\\
     &+\frac{8\epsilon_i\mu_i}{\pi^2}\left\lbrace J_i^{(s)}\left(\gamma-\frac{3}{2}\right)+\right.\\
     &\left.J_j^{(s)}\left[\gamma+\ln \left(\sqrt{\frac{\nu_i}{\nu_j}}\frac{|a_i|}{|a_j|}\right)-1-\beta_k \cot 2\beta_k \right]\right.\\
     &\left.+J_k^{(s)}\left[\gamma+\ln \left(\sqrt{\frac{\nu_i}{\nu_k}}\frac{|a_i|}{|a_k|}\right)-1-\beta_j \cot 2\beta_j \right]\right\rbrace\\
     &+\frac{6\epsilon_i^2}{\pi}\left[J_j^{(p)}\frac{\sin(4\beta_k)-4\beta_k}{\sin^2 (2\beta_k)}-J_k^{(p)}\frac{\sin(4\beta_j)-4\beta_j}{\sin^2 (2\beta_j)}\right].
\end{split}
\end{equation}
\end{subequations}
\section{THE GROUND STATE ENERGY OF THREE PARTICLES IN A PERIODIC BOX\label{sec:box}}

In this section,  we consider the ground state of 3 particles in a large periodic cubic box with side length $L$. Their wave function satisfies the periodic boundary condition:
\begin{equation}
\begin{split}
    &\Psi(\rr_1,\rr_2,\rr_3)=\Psi(\rr_1+\mathbf{L},\rr_2,\rr_3)\\
    &=\Psi(\rr_1,\rr_2+\mathbf{L},\rr_3)=\Psi(\rr_1,\rr_2,\rr_3+\mathbf{L}),
\end{split}
\end{equation}
where $\mathbf{L}=L(n_x\hat{\mathbf{e}}_x+n_y\hat{\mathbf{e}}_y+n_z\hat{\mathbf{e}}_z)$.
Here $\vect e_x$, $\vect e_y$, and $\vect e_z$ are unit vectors along the sides of the cube,
and $n_x$, $n_y$, and $n_z$ are integers.

Here we assume the two-body $s$-wave scattering lengths $a_1$, $a_2$, $a_3$ are 0, while the 3-body scattering hypervolume $D$ is not. If there are no two-body or three-body bound states, the three-body ground state wave function takes a simple form
\begin{equation}\label{Psiexpand}
\Psi(\rr_1,\rr_2,\rr_3)\approx 1-\frac{\sqrt{3}D}{8\pi^3 B^4}
\end{equation}
at $r_e\ll s_i\ll L$, where $r_e$ is the maximum range of two-body and three-body interactions.
The wave function should also satisfy the free Schr\"odinger equation
\begin{equation}
-\frac{\hbar^2}{2}\left( \nabla_1^2/m_1+\nabla_2^2/m_2+\nabla_3^2/m_3\right)\Psi=E\Psi \label{SE in box}
\end{equation}
if $s_1$, $s_2$ and $s_3$ are all greater than $r_e$.
Because the ground state has zero total momentum, $\Psi$ depends only on $(\mathbf{s}_{2},\mathbf{s}_{3})$,
%(where $\vect r_{21}=\vect r_2-\vect r_1$ and $\vect r_{31}=\vect r_3-\vect r_1$),
and can be written as $\Psi=\Psi(\mathbf{s}_{2},\mathbf{s}_{3})$. We multply both sides of \Eq{SE in box}
by $d^3s_{2} d^3s_{3}$ and integrate over $\vect s_{2}$ and $\vect s_{3}$ in the domain $B>B_0$
(where $r_e\ll B_0\ll L$): the right hand side yields approximately $EL^6$, and the left hand side can be computed by using Gauss's theorem and \Eq{Psiexpand}. We get
\beq
    E=\frac{\hbar^2 \widetilde{D}}{L^6},\label{E in cubic box}
\eeq
where
\beq\label{Dtilde and D}
\widetilde{D}\equiv\frac{\sqrt{3}(m_1m_2+m_2m_3+m_3m_1)^2}{9(m_1m_2m_3)^{3/2}\sqrt{m_1+m_2+m_3}}D.
\eeq
%\beq
%    \frac{\hbar^2D}{L^6}\frac{\sqrt{3}(m_1m_2+m_2m_3+m_3m_1)^2}{9(m_1m_2m_3)^{3/2}\sqrt{m_1+m_2+m_3}}\\
%    &\equiv \frac{\hbar^2 \tilde{D}}{L^6}.
%\eeq
The dimension of $\widetilde{D}$ is $[\mathrm{length}]^4/[\mathrm{mass}]$.

If there are three-body bound states but no two-body bound states,
Eqs.~\eqref{Psiexpand}, \eqref{SE in box}, \eqref{E in cubic box}, and \eqref{Dtilde and D}
are applicable to the lowest-energy three-body scattering state although this is no longer the three-body ground state.

If there are two-body bound states, \Eq{Psiexpand} would describe a metastable state rather than the ground state. This is analogous to real ultracold atomic gases which are after all not the true ground state of atoms (the true ground state of multiple atoms having attractive interactions is a tiny solid or liquid). One can tune the two-body interactions between ultracold atoms such that the scattering lengths are zero \cite{chin2010feshbach}, and then the three-body scattering hypervolumes will be among the dominant parameters for low-energy effective interactions.

\section{IMPLICATIONS FOR A DILUTE BOSE-BOSE MIXTURE\label{sec:bose-bose}}
We consider an interacting mixture of two Bose-Einstein condensed gases \cite{myatt1997production,ho1996binary,ao1998binary,hall1998dynamics,hall1998measurements,williams1999nonlinear,matthews1999vortices,busch1997stability,leggett2001bose,mudrich2002sympathetic}. If the
scattering lengths are tuned to zero
near a Feshbach resonance for cold atoms \cite{feshbach1958unified,chin2010feshbach} or if they are accidentally close to zero, or if the particles are near a low-energy three-body resonance, the interactions of the particles could be dominated by the three-body scattering hypervolumes.
%For certain atoms such as ${}^{88}\mathrm{Sr}$, the scattering lengths are accidentally tiny.
For a two-component Bose gas, consisting of bosons of types 1 and 2, there are four scattering hypervolumes, $D_{1}$, $D_{112}$, $D_{122}$, and $D_{2}$. Here $D_{1}$ is the intraspecies scattering hypervolume of three bosons of type 1,
$D_{112}$ is the scattering hypervolume of two bosons of type 1 and one boson of type 2,
$D_{122}$ is the scattering hypervolume of one boson of type 1 and two bosons of type 2,
and $D_{2}$ is the intraspecies scattering hypervolume of three bosons of type 2.

We consider $N_1$ bosons of type 1 and $N_2$ bosons of type 2 having vanishing or negligible
intraspecies and interspecies two-body scattering lengths ($a_{11}=a_{12}=a_{22}=0$)
in a large cubic box of side length $L$,
and impose the periodic boundary condition. Using \Eq{E in cubic box}, and assuming low enough densities
such that the total ground state energy $E$ may be approximated as a sum of the three-particle energies, we get
\begin{equation}
\begin{split}
E= \frac{\hbar^2}{L^6}\big(& C_{N_1}^{3}\widetilde{D}_{1}+C_{N_1}^{2}C_{N_2}^{1}\widetilde{D}_{112}\\
&+C_{N_1}^{1}C_{N_2}^{2}\widetilde{D}_{122}+C_{N_2}^{3}\widetilde{D}_{2}\big) ,
\end{split}
\end{equation}
where $C_{N}^{n}=N!/\big[n!(N-n)!\big]$. According to the general relations between $\widetilde{D}$ and $D$
in \Eq{Dtilde and D},
\begin{subequations}
\begin{align}
&\widetilde{D}_1=\frac{D_1}{m_1},\\
&\widetilde{D}_{112}=\frac{(m_1+2m_2)^2}{3m_1m_2\sqrt{3m_2(2m_1+m_2)}}D_{112},\\
&\widetilde{D}_{122}=\frac{(2m_1+m_2)^2}{3m_1m_2\sqrt{3m_1(m_1+2m_2)}}D_{122},\\
&\widetilde{D}_{2}=\frac{D_2}{m_2},
\end{align}
\end{subequations}
where $m_1$ is the mass of each boson of type 1, and $m_2$ is the mass of each boson of type 2.
In the thermodynamic limit, in which $N_1$, $N_2$, $L\rightarrow \infty$ while the densities $n_1=N_1/L^3$ and $n_2=N_2/L^3$ are fixed, we get
\begin{equation}
\begin{split}
\frac{E}{\hbar^2\Omega}&=\frac{1}{6} \widetilde{D}_{1}n_1^3+\frac{1}{6} \widetilde{D}_{2}n_2^3\\
&+\frac{1}{2}\widetilde{D}_{112}n_1^2 n_2+\frac{1}{2}\widetilde{D}_{122}n_1n_2^2,
\end{split}\label{Energy of two-BEC}
\end{equation}
%or
%\begin{equation}
%\begin{split}
%\frac{\mathcal{E}}{\hbar^2}&=\frac{1}{6}\rho_1^3 \tilde{D}_{111}+\frac{1}{2} \rho_1^2\rho_2 \tilde{D}_{112}\\
%&+\frac{1}{2}\rho_1\rho_2^2 \tilde{D}_{122}+\frac{1}{6}\rho_2^3 \tilde{D}_{222},
%\end{split}\label{energy density}
%\end{equation}
where $\Omega=L^3$ is the volume of the system. The energy can also be calculated using the Effective Field Theory (EFT) \cite{braaten1999quantum}.

Taking the partial derivative of the energy with respect to the densities $n_1$ or $n_2$, we get the chemical potentials $\mu_1$ and $\mu_2$. Further adding the kinetic energy operators and the external potentials, we find that
the two-component Bose-Einstein condensate (BEC) can be described by the following coupled Gross-Pitaevskii equations \cite{gross1961structure,pitaevskii1961vortex},
\begin{widetext}
\begin{subequations}
\begin{align}
\mathrm{ i }\hbar\frac{\partial}{\partial t}\Psi_1
=\Big[ &-\frac{\hbar^2\nabla^2}{2m_1}+V_1(\vect r,t)+\frac{\hbar^2}{2}\widetilde{D}_{1}|\Psi_1|^4+\hbar^2\widetilde{D}_{112}|\Psi_1|^2|\Psi_2|^2+\frac{\hbar^2}{2}\widetilde{D}_{122}|\Psi_2|^4 \Big]\Psi_1,\\
\mathrm{ i }\hbar\frac{\partial}{\partial t}\Psi_2=\Big[ &-\frac{\hbar^2\nabla^2}{2m_2}+V_2(\vect r,t)+\frac{\hbar^2}{2}\widetilde{D}_{112}|\Psi_1|^4+\hbar^2\widetilde{D}_{122}|\Psi_1|^2|\Psi_2|^2+\frac{\hbar^2}{2}\widetilde{D}_{2}|\Psi_2|^4 \Big] \Psi_2,
\end{align}
\end{subequations}
\end{widetext}
where $\Psi_1=\Psi_1(\vect r,t)$ and $\Psi_2=\Psi_2(\vect r,t)$ are the macroscopic wave functions
whose norm-squares are the densities $n_1$ and $n_2$ respectively, and $V_1(\vect r,t)$ and $V_2(\vect r,t)$
are the external potentials experienced by the bosons of types 1 and 2 respectively. The terms containing $|\Psi_1|^4$, $|\Psi_1|^2|\Psi_2|^2$ or $|\Psi_2|^4$ in the above two equations generalize the three-body coupling term in the Gross-Pitaevskii equation for a single-component BEC \cite{gammal2000atomic,kohler2002three}.
%Rather than defining the three-body coupling constants phenomenalogically,
We have related the three-body coupling constants to the wave functions for the zero-energy collisions of three particles, facilitating numerical determinations of these constants
for any given microscopic interactions.

In a two-component BEC, if the two-body scattering lenghts $a_{ij}$ are not zero, the system is stable when $g_{11}>0$, $g_{22}>0$, and $|g_{12}|<\sqrt{g_{11}g_{22}}$ \cite{pethick2008bose,pitaevskii2016bose}, where $g_{ij}=2\pi\hbar^2 a_{ij}/\mu_{ij}$ is the two-body coupling constant, $\mu_{ij}$ is the reduced mass of $m_i$ and $m_j$. The first and second conditions ensure stability against collapse when only one component exists. The third condition ensures the two species are mixed together, rather than phase separated  \cite{myatt1997production,timmermans1998phase,ho1996binary,lee2016phase,tojo2010controlling,wen2012controlling}.

Now we derive the conditions for the stability of the two-component BEC with only three-body scattering hypervolumes and
negligible two-body scattering lengths. The zero-temperature state energy of a homogeneous gaseous mixture of the two components is given by \Eq{Energy of two-BEC}.
The mixture should be dynamically stable against local density fluctuations \cite{pethick2008bose} if
\beq
\partial^2 E_{\mathrm{mix}}/\partial n_i^2>0
\eeq
and
\begin{equation}
    \left(\frac{\partial^2 E}{\partial n_1^2} \right)\left(\frac{\partial^2 E}{\partial n_2^2} \right)>\left(\frac{\partial^2 E}{\partial n_1 \partial n_2}\right)^2.
\end{equation}
Substituting \Eq{Energy of two-BEC} into the above inequalities, we get
\begin{equation}
    \begin{split}
    &\widetilde{D}_1 n_1+\widetilde{D}_{112}n_2>0,\\
    &\widetilde{D}_2 n_2+\widetilde{D}_{122}n_1>0,
    \end{split}
\end{equation}
\begin{equation}
\begin{split}
     &\left(\widetilde{D}_1 n_1+\widetilde{D}_{112}n_2\right)\left(\widetilde{D}_2 n_2+\widetilde{D}_{122}n_1\right)\\
    &>\left(\widetilde{D}_{112}n_1+\widetilde{D}_{122}n_2\right)^2.
\end{split}
\end{equation}
%where $\widetilde{D}_1=\widetilde{D}_{111}$, $\widetilde{D}_2=\widetilde{D}_{222}$.

If the scattering hypervolumes become complex (with negative imaginary parts \cite{zhu2017three}), the energy in \Eq{Energy of two-BEC} gains a negative imaginary part, indicating the decaying of the BEC. Within a short time $\Delta t$, the probability that no recombination occurs is $\mathrm{exp}(-2|\mathrm{Im} E|\Delta t/\hbar)\simeq 1-2|\mathrm{Im} E|\Delta t/\hbar$. Then the probability for one recombination is $2|\mathrm{Im} E|\Delta t/\hbar$. If
the BEC is contained in a shallow trap (whose depth is small compared to the energy released in a three-body recombination event),
after each recombination event, three atoms escape from the trap. This leads to the decay rates of the atomic densities
within the trap:
\begin{equation}
\begin{split}
    &\frac{1}{\hbar}\frac{\mathrm{d}n_1}{\mathrm{d}t}=-|\mathrm{Im}\widetilde{D}_1|n_1^3-2|\mathrm{Im}\widetilde{D}_{112}| n_1^2 n_2-|\mathrm{Im}\widetilde{D}_{122}| n_1 n_2^2,\\
    &\frac{1}{\hbar}\frac{\mathrm{d}n_2}{\mathrm{d}t}=-|\mathrm{Im}\widetilde{D}_2|n_2^3-2|\mathrm{Im}\widetilde{D}_{122}| n_1 n_2^2-|\mathrm{Im}\widetilde{D}_{112}| n_1^2 n_2.
\end{split}
\end{equation}

%\section{Three-particle scattering amplitude}
%In this section, we compute the three-body scattering amplitude at low energy limit with the $s$-wave scattering length being 0 but the three-body parameter $D\neq 0$.

%We consider 3 particles with incoming momenta $\mathbf{b}_1$, $\mathbf{b}_2$, $\mathbf{b}_3$, and the energy
%\begin{equation}
%E=\frac{b_1^2}{2m_1}+\frac{b_2^2}{2m_2}+\frac{b_3^2}{2m_3}
%\end{equation}
%is conserved.

%One can extract the three-particle T-matrix elements from the wavefunction:
%\begin{equation}
%\begin{split}
%\psi_{\mathbf{q}_1\mathbf{q}_2\mathbf{q}_3}=&(2\pi)^6\delta(\mathbf{q}_1-\mathbf{b}_1)\delta(\mathbf{q}_2-\mathbf{b}_2)\\
%&+G_{q_1q_2q_3}^{E}T\left( \mathbf{q}_1\mathbf{q}_2\mathbf{q}_3;\mathbf{b}_1\mathbf{b}_2\mathbf{b}_3\right)\\ 
%+&\Big( \mathrm{terms \ that \ are \ regular \ at\ }\\
%&\frac{q_1^2}{2m_1}+\frac{q_2^2}{2m_2}+ \frac{q_3^2}{2m_3}=E  \Big) 
%\end{split}\label{scatter wavefunction with T matrix}
%\end{equation}
%where $G_{q_1q_2q_3}^{E}$ is 
%\begin{equation}
%G_{q_1q_2q_3}^{E}=\left( \frac{q_1^2}{2m_1}+\frac{q_2^2}{2m_2}+ %\frac{q_3^2}{2m_3}-E-\mathrm{i}\epsilon\right)  ^{-1}.
%\end{equation}

%In the zero energy limit, compare \ref{scatter wavefunction with T matrix} and \ref{wavefunction q1q2q3}, we get 
%\begin{equation}
%T\left( \mathbf{q}_1\mathbf{q}_2\mathbf{q}_3;\mathbf{b}_1\mathbf{b}_2\mathbf{b}_3\right)=-\tilde{D}.
%\end{equation}

%It's just a constant and is not dependent on the momenta, just like two-body $s$-wave scattering in the low energy limit.

\section{Summary}
We studied the wave function for the collision of three particles of unequal masses with short-range interactions at zero incoming kinetic energy and zero orbital angular momentum. We derived the asymptotic expansions of such
a wave function when two particles are held at a fixed distance and the third particle is far away from the two,
or when all three particles are far away from each other. From these expansions we defined the three-body scattering hypervolume for the three particles. This generalizes the definition of three-body scattering hypervolume for identical bosons in Ref.~\cite{tan2008three}. We then computed the ground state energy of three particles of unequal masses with short-range interactions in a large cubic box, assuming vanishing two-body scattering lengths. This result enabled us to compute the zero-temperature energy of a dilute two-component Bose-Einstein condensate (BEC) having vanishing or negligible two-body scattering lengths, to write down the corresponding Gross-Pitaevskii equation for such a BEC
in some external potentials, to derive conditions for the stability of the mixture, and to find the decay
rates of particle densities due to three-body recombination events.

\begin{acknowledgments}
This work was supported by the National Key R\&D Program of China (Grants No.~2019YFA0308403).
\end{acknowledgments}


\appendix

\section*{Appendix: \label{app-1}Procedure for determining the 1-1-1 expansion and 2-1 expansion}
If $s_i\ll R_i$, we can expand $T^{(-p)}$ as
\begin{equation}
T^{(-p)}=\sum_{n} t_i^{(n,-p-n)},
\eeq
where $t_i^{(n,m)}$ scales like $R_i^n s_i^m$ (with a possible extra factor that scales like a polynomial of
$\ln R_i$).
If $s_i\gg r_e$, we can expand $S_i^{(-q)}$ as
\beq
S_{i}^{(-q)}=\sum_{m} t_i^{(-q,m)}.
\end{equation}
Because the three-body wave function $\Phi^{(3)}$ may be expanded as $\sum_p T^{(-p)}$ at $B\to\infty$,
and may also be expanded as $\sum_q S^{(-q)}$ at $R_i\to\infty$, the $t_i^{(n,m)}$ in the above two expansions
should be the same. In fact the wave function has a double expansion $\Phi^{(3)}=\sum_{n,m}t_i^{(n,m)}$ in the region $r_e\ll s_i\ll R_i$.

We choose the overall amplitude of $\Phi^{(3)}$ such that $T^{(0)}=1$. Therefore
\begin{equation}
\begin{split}
    &t_i^{(0,0)}=1,\\
    &t_i^{(-1,1)}=0,\\
    &t_i^{(-2,2)}=0,\\
    &\cdots
\end{split}\label{t_00}
\end{equation}
From 
\begin{equation}
    \hat{H}_i S_i^{(0)}=0,
\end{equation}
we deduce that $S_i^{(0)}$ takes the form
\begin{equation}
    S_i^{(0)}=\sum_l c_{l} \phi_{i,\hat{\vect{R}}_i}^{(l)}(\vect{s}_i).
\end{equation}
Using the expansion $S_i^{(0)}=t_i^{(0,0)}+t_i^{(0,-1)}+...$ at $s_i\gg r_e$, we find that 
here the coefficient $c_0=1$ but $c_l=0$ for $l\ge1$. So
\beq
S_i^{(0)}=\phi_i(\vect s_i).
\eeq
If $s_i>r_e$ we get
\begin{equation}
    S_i^{(0)}=1-\frac{a_i}{s_i}.
\end{equation}
This leads to
\beq
t_i^{(0,-1)}=-\frac{a_i}{s_i},
\eeq
and it will contribute to $T^{(-1)}$.

$T^{(-1)}$ should satisfy \Eq{free Schrodinger} outside the interaction range, and $T^{(-1)}=t_i^{(0,-1)}+t_i^{(-1,0)}+t_i^{(-2,1)}+...$ if $s_i\ll R_i$. From these conditions we can determine $T^{(-1)}$:
\begin{equation}
    T^{(-1)}=\sum_{i=1}^{3} -\frac{a_i}{s_i}.
\end{equation}
Expanding $T^{(-1)}$ at $s_i\ll R_i$, we get 
\begin{equation}
\begin{split}
    &t_i^{(-1,0)}=-\frac{a_j+a_k}{R_i},\\
    &t_i^{(-2,1)}=(\eta_{jk}a_j-\eta_{kj}a_k)\frac{s_i}{R_i^2}P_1 (\hat{\vect{R}}_i\cdot\hat{\vect{s}}_i),\\
    &t_i^{(-3,2)}=-(\eta_{jk}^2a_j+\eta_{kj}^2a_k)\frac{s_i^2}{R_i^3}P_2 (\hat{\vect{R}}_i\cdot\hat{\vect{s}}_i),\\
    &t_i^{(-4,3)}=(\eta_{jk}^3a_j-\eta_{kj}^3a_k)\frac{s_i^3}{R_i^4}P_3 (\hat{\vect{R}}_i\cdot\hat{\vect{s}}_i),\\
\end{split}
\end{equation}
and so on.
From the expansion $S_i^{(-1)}=t_i^{(-1,1)}+t_i^{(-1,0)}+t_i^{(-1,-1)}+...$ at $s_i\gg r_e$, and
\begin{equation}
    \hat{H}_i S_i^{(-1)}=0,
\end{equation}
%So $S_i^{(-1)}$ takes the form:
%\begin{equation}
%    S_i^{(-1)}=\frac{1}{R_i}\sum_l B_{l} \phi_{i,\hat{\vect{R}}_i}^{(l)}(\vect{s}_i)
%\end{equation}
%Here $l\leq1$, otherwise $T^{(p)}$ ($p>0$) will exist. Then
%\begin{equation}
%    S_i^{(-1)}=\frac{1}{R_i} B_{0} \phi_i(\vect{s}_i)+\frac{1}{R_i} B_{1}\phi_{i,\hat{\vect{R}}_i}^{(p)}(\vect{s}_i).
%\end{equation}
%Since $t^{(-1,1)}=0$, $t^{(-1,0)}=-(a_j+a_k)/R_i$,
we find
\begin{equation}
     S_i^{(-1)}=-\frac{a_j+a_k}{R_i}\phi_i(\vect{s}_i).
\end{equation}
This leads to
\beq
t_i^{(-1,-1)}=\frac{a_i(a_j+a_k)}{R_i s_i},
\eeq
and it will contribute to $T^{(-2)}$.

Repeating this procedure, we can successively determine $T^{(-2)}$, $S_i^{(-2)}$, ..., $T^{(-4)}$, and $S_i^{(-4)}$. In this way we computed the three-body wave function order by order, and finally arrived at the 111-expansion \Eq{1-1-1 expansion} and the 21-expansion \Eq{2-1 expansion}.

%\nocite{*}

\bibliography{ref}% Produces the bibliography via BibTeX.

\end{document}